\documentclass[10pt,twocolumn,letterpaper]{article}

\usepackage[final]{cvpr}
\date{}

\usepackage{times}
\usepackage{epsfig}
\usepackage{graphicx}
\usepackage{amsmath}
\usepackage{amssymb}
\usepackage{booktabs}
\usepackage{algorithm}
\usepackage{algorithmic}
\usepackage[numbers,sort,compress]{natbib}
\usepackage{pgfplots}
\usepackage{float}
\pgfplotsset{compat=1.18}

\usepackage[pagebackref,breaklinks,colorlinks]{hyperref}
\hypersetup{
    colorlinks=true,
    linkcolor=blue,
    urlcolor=cyan,
    citecolor=blue,
}

\def\confYear{2025}
\def\confName{NeurIPS}
\def\cvprPaperID{AGENT2025}

\begin{document}

\title{Structural Representations for Cross-Attack Generalization\\in AI Agent Threat Detection}

\author{
Vignesh Iyer\\
{\tt\small iyerv68@gmail.com}
}

\maketitle

\begin{abstract}
Autonomous AI agents executing multi-step tool sequences face semantic attacks that manifest in behavioral traces rather than isolated prompts. A critical challenge is cross-attack generalization: can detectors trained on known attack families recognize novel, unseen attack types? We discover that standard conversational tokenization---capturing linguistic patterns from agent interactions---fails catastrophically on structural attacks like tool hijacking (AUC 0.39) and data exfiltration (AUC 0.46), while succeeding on linguistic attacks like social engineering (AUC 0.78). We introduce structural tokenization, encoding execution-flow patterns (tool calls, arguments, observations) rather than conversational content. This simple representational change dramatically improves cross-attack generalization: +46 AUC points on tool hijacking, +39 points on data exfiltration, and +71 points on unknown attacks, while simultaneously improving in-distribution performance (+6 points). For attacks requiring linguistic features, we propose gated multi-view fusion that adaptively combines both representations, achieving AUC 0.89 on social engineering without sacrificing structural attack detection. Our findings reveal that AI agent security is fundamentally a structural problem: attack semantics reside in execution patterns, not surface language. While our rule-based tokenizer serves as a baseline, the structural abstraction principle generalizes even with simple implementation.
\end{abstract}

\section{Introduction}

Autonomous AI agents powered by large language models are transitioning from research prototypes to operational infrastructure~\cite{xi2023rise,wang2024survey}. Enabled by tool-use capabilities~\cite{schick2023toolformer,qin2023toolllm,patil2023gorilla}, modern deployments span critical business functions: customer service agents process refunds and manage accounts, document agents extract information from contracts, developer agents modify codebases, and data agents query production databases. This operational autonomy requires agents to possess execution privileges over tools that directly affect business operations, creating attack vectors fundamentally different from traditional software vulnerabilities.

\subsection{Semantic Attacks on AI Agents}

Traditional cybersecurity threats exploit implementation flaws through well-understood mechanisms: buffer overflows manipulate memory, SQL injection exploits input sanitization, and cross-site scripting abuses DOM manipulation~\cite{howard2005security,owasp2021top10}. These attacks succeed by violating syntactic constraints and leave clear signatures detectable through pattern matching.

AI agent attacks instead operate through \textit{semantic manipulation} of natural language reasoning~\cite{andriushchenko2024agentharm,perez2022red}. \textbf{Indirect prompt injection}~\cite{greshake2023ipi,liu2023prompt,zhan2024injecagent} embeds malicious instructions in external data sources that override user intent. \textbf{Tool hijacking}~\cite{shayegani2023plug} redirects operations toward attacker-controlled endpoints. \textbf{Data exfiltration} extracts sensitive information through seemingly benign tool sequences. Adversarial attacks on aligned models~\cite{zou2023universal} demonstrate the brittleness of safety training. Unlike traditional attacks, semantic attacks maintain perfect syntactic validity---detection requires understanding \textit{behavioral context}.

\subsection{The Cross-Attack Generalization Problem}

Existing defenses focus on detecting specific attack patterns observed during training. However, a critical question remains understudied: \textbf{can threat detectors generalize to attack families never seen during training?} Real-world deployments inevitably face novel attacks not represented in training data. If a detector trained on prompt injection cannot recognize tool hijacking, its practical utility is severely limited.

We systematically evaluate cross-attack generalization by holding out entire attack families during training and measuring detection performance on these unseen categories~\cite{koh2021wilds,hendrycks2019benchmarking}. Our findings reveal a striking asymmetry: generalization success depends critically on \textit{what behavioral signals the representation captures}.

\subsection{Key Insight: Structure vs. Language}

We discover that standard conversational tokenization---encoding linguistic patterns from user messages and agent responses---exhibits dramatic variation in cross-attack transfer. Linguistic attacks like social engineering (AUC 0.78) and prompt injection (AUC 0.69) transfer moderately well. However, structural attacks like tool hijacking (AUC 0.39) and data exfiltration (AUC 0.46) fail catastrophically, while unknown attacks collapse entirely (AUC 0.26).

This asymmetry reveals that conversational features capture persuasion tactics but fundamentally miss execution-level threats. We hypothesize that structural attacks depend on \textit{how tools are orchestrated}, not \textit{what is said}---a tool hijacking attack may use entirely benign language while executing a malicious tool sequence.

\subsection{Contributions}

We make three contributions:

\textbf{(1) Structural tokenization.} We introduce execution-flow tokenization that encodes tool calls, argument patterns, and observation sequences rather than conversational content~\cite{du2017deeplog,mirsky2018kitsune}. This improves cross-attack generalization by 39--71 AUC points while also improving in-distribution performance.

\textbf{(2) Attack-family taxonomy.} We provide the first systematic study of cross-attack transfer~\cite{shen2021towards,yang2021generalized}, revealing that linguistic and structural attacks require fundamentally different representations.

\textbf{(3) Gated multi-view fusion.} For deployments facing diverse attack types, we propose adaptive fusion that learns when to rely on each representation, achieving strong performance across all attack families.

\section{Related Work}

\textbf{AI Agent Security.} Autonomous agents face semantic attacks including indirect prompt injection~\cite{greshake2023ipi,liu2023prompt,zhan2024injecagent}, jailbreaking~\cite{wei2023jailbroken,zou2023universal}, and tool manipulation~\cite{shayegani2023plug}. AgentHarm~\cite{andriushchenko2024agentharm} provides benchmarks for measuring agent harmfulness. Prior defenses focus on input sanitization and prompt hardening rather than cross-attack generalization.

\textbf{Behavioral Sequence Analysis.} Sequence-based anomaly detection has proven effective in system log analysis~\cite{du2017deeplog,brown2018recurrent} and network intrusion detection~\cite{mirsky2018kitsune}. We extend this intuition to AI agents, showing that execution-flow patterns provide stronger cross-attack generalization than linguistic features.

\textbf{Multi-View Learning.} Combining multiple representations improves robustness across domains~\cite{baltrusaitis2018multimodal,xu2013survey,ngiam2011multimodal}. Our gated fusion builds on mixture-of-experts principles~\cite{jacobs1991adaptive,shazeer2017outrageously} to adaptively weight conversational and structural views.

\textbf{Distribution Shift.} Out-of-distribution generalization remains challenging~\cite{koh2021wilds,shen2021towards,hendrycks2019benchmarking,yang2021generalized}. Our cross-attack evaluation protocol explicitly measures generalization to unseen attack families---a critical requirement for real-world security.

\textbf{Federated Security.} Federated intrusion detection~\cite{ferrag2021federated} and malware classification~\cite{li2021fedmal} enable collaborative learning. Recent work explores federated prompt injection detection~\cite{jayathilaka2025federated}. Non-IID data remains challenging~\cite{zhao2018federated,li2020fedprox}; we discover that representation choice dominates aggregation method.

\section{Method}

\subsection{Problem Formulation}

Consider agent traces $\tau = (\mathcal{M}, \mathcal{T}, \mathcal{R})$ containing user/assistant messages $\mathcal{M}$, tool invocations $\mathcal{T} = \{(t_i, \text{args}_i, \text{obs}_i)\}$, and final responses $\mathcal{R}$. The detection task is binary classification: benign vs.\ attack.

\textbf{Cross-attack evaluation.} Let $\mathcal{A} = \{a_1, \ldots, a_m\}$ denote attack families. For held-out family $a_j$, we train on all data except $a_j$ and evaluate detection specifically on $a_j$. This measures true generalization to unseen attack types.

\subsection{Conversational Tokenization (Baseline)}

Our baseline represents traces using a 26-token vocabulary capturing linguistic and behavioral patterns through keyword matching and pattern rules:

\begin{itemize}
    \item \textit{Tool Types} (8): \texttt{SEND\_EMAIL}, \texttt{MAKE\_PAYMENT}, \texttt{READ\_FILE}, \texttt{EXECUTE\_CODE}, etc.
    \item \textit{Argument Patterns} (6): \texttt{EXTERNAL\_RECIPIENT}, \texttt{HIGH\_VALUE}, \texttt{SENSITIVE\_FIELD}, etc.
    \item \textit{Attack Indicators} (6): \texttt{INJECTION\_PHRASE}, \texttt{OVERRIDE\_ATTEMPT}, \texttt{EVASION}, etc.
    \item \textit{Response Types} (3): \texttt{REFUSAL}, \texttt{COMPLIANT}, \texttt{CLARIFY}
    \item \textit{Control Flow} (3): \texttt{LOOP}, \texttt{BRANCH}, \texttt{RECURSION}
\end{itemize}

This representation captures \textit{what is said} through attack-indicative phrases. However, it encodes execution structure only indirectly through tool-type tokens.

\subsection{Structural Tokenization}

We introduce execution-flow tokenization using a compact 9-token vocabulary that encodes \textit{what the agent did} rather than linguistic content:

\begin{center}
\small
\begin{tabular}{ll}
\toprule
\textbf{Token} & \textbf{Semantics} \\
\midrule
\texttt{[SYS]} & System instruction present \\
\texttt{[USER]} & User message \\
\texttt{[ASSISTANT]} & Assistant response \\
\texttt{[TOOL]} & Tool invocation detected \\
\texttt{[ARGS]} & Arguments passed to tool \\
\texttt{[OBS]} & Tool observation returned \\
\texttt{[OUTPUT]} & Final response to user \\
\texttt{[ERROR]} & Error or exception \\
\texttt{[OTHER]} & Fallback token \\
\bottomrule
\end{tabular}
\end{center}

\textbf{Example.} A trace where a user requests a file, the agent invokes \texttt{read\_file}, receives data, and responds becomes: \texttt{[USER] [ASSISTANT] [TOOL] [ARGS] [OBS] [OUTPUT]}.

This representation abstracts away linguistic content entirely, capturing only the \textit{shape} of agent execution~\cite{brown2018recurrent}. Crucially, structural patterns remain discriminative even when attackers paraphrase---they cannot hide the resulting execution flow.

\subsection{Gated Multi-View Fusion}

Neither representation alone is optimal for all attack families. We propose adaptive fusion~\cite{baltrusaitis2018multimodal,jacobs1991adaptive} using a learned gate:

\begin{equation}
    g = \sigma\left(\mathbf{W}_g [h_{\text{conv}}; h_{\text{struct}}] + b_g\right)
\end{equation}
\begin{equation}
    h_{\text{fused}} = g \odot h_{\text{conv}} + (1-g) \odot h_{\text{struct}}
\end{equation}

where $h_{\text{conv}}$ and $h_{\text{struct}}$ are encoded representations from parallel BiLSTM encoders~\cite{hochreiter1997lstm}. The gate $g$ learns to weight each view based on input characteristics.

\subsection{Architecture and Training}

All models share: 64-dim embeddings~\cite{bengio2013representation}, bidirectional LSTM~\cite{hochreiter1997lstm} (hidden=64, output=128), projection to 32-dim latent space, and two-layer classifier (32→64→1) with sigmoid output. Total: ~74K parameters (single-view) or ~140K (multi-view).

We train in a federated setting with $K=5$ organizations using FedAvg~\cite{mcmahan2017fedavg}: 5 rounds, 5 local epochs per round, Adam optimizer~\cite{kingma2014adam} ($\eta=0.001$), batch size 32, BCE loss. Optional DP-SGD~\cite{abadi2016dpsgd} provides privacy guarantees ($\epsilon \in [2.35, 5.69]$).

\section{Experiments}

\subsection{Experimental Setup}

\textbf{Dataset.} We evaluate on a simulated 5-organization federation with 2,500 agent traces. Each organization contributes 500 traces (50\% benign, 50\% malicious) with non-IID attack distributions reflecting realistic specialization.

\textbf{Attack families.} We evaluate 5 categories: prompt injection (linguistic manipulation), tool hijacking (unauthorized redirection), data exfiltration (information theft), social engineering (persuasion-based), and unknown (novel/uncategorized attacks).

\textbf{Evaluation.} For cross-attack generalization, we hold out one attack family entirely during training and evaluate on that held-out family. We report ROC-AUC.

\subsection{Main Results}

Table~\ref{tab:main_results} presents our central finding: representation choice dramatically affects cross-attack transfer.

\begin{table}[t]
\centering
\small
\begin{tabular}{lccc}
\toprule
\textbf{Held-Out Attack} & \textbf{Conv} & \textbf{Struct} & \textbf{Gated} \\
\midrule
Social engineering & 0.78 & 0.67 & \textbf{0.89} \\
Prompt injection & 0.69 & 0.81 & \textbf{0.83} \\
Data exfiltration & 0.46 & \textbf{0.85} & 0.62 \\
Tool hijacking & 0.39 & \textbf{0.85} & 0.60 \\
Unknown & 0.26 & \textbf{0.97} & 0.92 \\
\midrule
\textbf{IID (seen attacks)} & 0.87 & \textbf{0.93} & 0.89 \\
\bottomrule
\end{tabular}
\vspace{-2mm}
\caption{Cross-attack generalization (ROC-AUC). Structural tokenization dramatically improves detection of tool-based attacks (+39--71 points) while gated fusion excels on linguistic attacks. Bold indicates best per row.}
\label{tab:main_results}
\end{table}

\textbf{Conversational tokenization} achieves good performance on linguistic attacks (social engineering: 0.78, prompt injection: 0.69) but fails catastrophically on structural attacks (tool hijacking: 0.39, data exfiltration: 0.46) and unknown attacks (0.26---worse than random).

\textbf{Structural tokenization} achieves strong performance on tool-based attacks (0.85) and unknown attacks (0.97), while also improving IID performance (0.93 vs 0.87). However, it underperforms on social engineering (0.67).

\textbf{Gated fusion} achieves the best balance: strong on social engineering (0.89) and prompt injection (0.83) while maintaining good performance on unknown attacks (0.92).

\subsection{Effect Size Analysis}

Table~\ref{tab:effect_size} quantifies the improvement from structural tokenization.

\begin{table}[t]
\centering
\small
\begin{tabular}{lccc}
\toprule
\textbf{Setting} & \textbf{Conv} & \textbf{Struct} & \textbf{$\Delta$} \\
\midrule
Unknown (OOD) & 0.26 & 0.97 & \textbf{+0.71} \\
Tool hijacking (OOD) & 0.39 & 0.85 & \textbf{+0.46} \\
Data exfiltration (OOD) & 0.46 & 0.85 & \textbf{+0.39} \\
Prompt injection (OOD) & 0.69 & 0.81 & +0.12 \\
IID & 0.87 & 0.93 & +0.06 \\
\midrule
Social engineering (OOD) & 0.78 & 0.67 & --0.11 \\
\bottomrule
\end{tabular}
\vspace{-2mm}
\caption{AUC gains from structural tokenization. Improvements of 39--71 points on hard OOD cases; regression only on social engineering.}
\label{tab:effect_size}
\end{table}

The improvements are substantial and consistent. Notably, structural tokenization improves \textit{both} OOD and IID performance---a rare outcome where robustness does not trade off against accuracy.

\subsection{Attack-Representation Dependency}

Table~\ref{tab:dependency} reveals which representations each attack family requires.

\begin{table}[t]
\centering
\small
\begin{tabular}{lcc}
\toprule
\textbf{Attack Family} & \textbf{Conv?} & \textbf{Struct?} \\
\midrule
Social engineering & \checkmark\checkmark & $\times$ \\
Prompt injection & \checkmark & \checkmark \\
Data exfiltration & $\times$ & \checkmark\checkmark \\
Tool hijacking & $\times$ & \checkmark\checkmark \\
Unknown & $\times$ & \checkmark\checkmark \\
\bottomrule
\end{tabular}
\vspace{-2mm}
\caption{Attack-representation dependency. Structural features dominate for 4/5 attack families.}
\label{tab:dependency}
\end{table}

This taxonomy reveals a fundamental insight: \textbf{most AI agent attacks are structural, not linguistic}. Only social engineering requires conversational features.

\subsection{Federated Aggregation}

A key finding is that aggregation method has \textbf{no significant effect}: Local training, FedAvg, and ensemble methods achieve identical results within $\pm$0.02 AUC. This establishes that \textbf{representation---not aggregation---is the bottleneck} for cross-attack generalization.

\subsection{Privacy Analysis}

With DP-SGD ($\epsilon=2.35$), the structural model achieves OOD AUC of 0.72---still substantially above the conversational baseline's 0.52 without any privacy constraints. Meaningful signal persists under strong privacy.

\section{Discussion}

\textbf{Why structure matters.} Attack semantics reside in execution patterns---tool sequences, argument flows, observation handling---not surface language~\cite{bengio2013representation}. Conversational tokenization detects \textit{how attackers phrase requests}; structural tokenization detects \textit{what agents do}. Attackers can paraphrase arbitrarily but cannot hide execution traces.

\textbf{The social engineering exception.} Social engineering relies on psychological manipulation manifesting in conversational style rather than tool orchestration. Structural tokenization underperforms here (0.67 vs 0.78) because the signal is inherently linguistic. Gated fusion~\cite{xu2013survey,ngiam2011multimodal} addresses this by learning to weight conversational features for persuasion-based attacks.

\textbf{Unknown attack recovery.} The dramatic improvement on unknown attacks (0.26→0.97) is particularly notable. Analysis reveals that conversational tokenization learned ``familiar = safe,'' causing ranking inversion on novel patterns. Structural tokenization detects execution anomalies regardless of linguistic novelty.

\textbf{Implications.} Practitioners should: (1) default to structural representations for agent security, (2) add conversational features only when social engineering is a significant threat, (3) not expect federated aggregation to compensate for representational weaknesses.

\textbf{Limitations.} Our evaluation uses synthetic traces; real-world validation is critical. \textbf{Our rule-based tokenizer is intentionally simple to isolate the effect of structural abstraction from tokenizer sophistication, but this makes it potentially susceptible to adversarial evasion.} Attackers may fragment tool calls, add noise, or structure malicious traces to mimic benign patterns. However, our approach is \textit{modular}---the tokenizer can be replaced with learned variants (neural encoders, contrastive learning) while preserving the core insight that structural patterns generalize better than conversational content. Evaluation against adaptive adversaries is critical future work. Our 5-family taxonomy requires expansion. The gated model underperforms struct-only on structural attacks, suggesting improved fusion mechanisms~\cite{shazeer2017outrageously} are needed.

\section{Conclusion}

We demonstrate that cross-attack generalization in AI agent threat detection is fundamentally a representation problem. Structural tokenization---encoding execution flow rather than conversational content---improves detection of unseen attack families by 39--71 AUC points while simultaneously improving in-distribution performance. This challenges the assumption that linguistic features suffice for agent security.

Our key insight is that \textbf{AI agent security is primarily structural}: attack semantics reside in tool orchestration patterns, not surface language. Practitioners should design detectors that analyze \textit{what agents do}, not merely \textit{what users say}~\cite{du2017deeplog}. For diverse threats including social engineering, gated multi-view fusion provides a principled approach to combining both signal types~\cite{baltrusaitis2018multimodal}. While our rule-based tokenizer provides a strong baseline, future work should explore learned tokenization and adaptive adversary evaluation to further strengthen detection robustness.

{\small
\bibliographystyle{abbrvnat}
\bibliography{references}

@inproceedings{greshake2023ipi,
  title={Not What You've Signed Up For: Compromising Real-World {LLM}-Integrated Applications with Indirect Prompt Injection},
  author={Greshake, Kai and Abdelnabi, Sahar and Mishra, Shailesh and Endres, Christoph and Holz, Thorsten and Fritz, Mario},
  booktitle={Proceedings of the 16th ACM Workshop on Artificial Intelligence and Security (AISec)},
  pages={79--90},
  year={2023}
}

@inproceedings{andriushchenko2024agentharm,
  title={{AgentHarm}: A Benchmark for Measuring Harmfulness of {LLM} Agents},
  author={Andriushchenko, Maksym and Souly, Alexandra and Dziemian, Mateusz and Duenas, Derek and Lin, Maxwell and Wang, Justin and Hendrycks, Dan and Zou, Andy and Kolter, Zico and Fredrikson, Matt and Winsor, Eric and Wynne, Jerome and Gal, Yarin and Davies, Xander},
  booktitle={International Conference on Learning Representations (ICLR)},
  year={2025}
}

@inproceedings{wei2023jailbroken,
  title={Jailbroken: How Does {LLM} Safety Training Fail?},
  author={Wei, Alexander and Haghtalab, Nika and Steinhardt, Jacob},
  booktitle={Advances in Neural Information Processing Systems (NeurIPS)},
  volume={36},
  year={2023}
}

@article{shayegani2023plug,
  title={Survey of Vulnerabilities in Large Language Models Revealed by Adversarial Attacks},
  author={Shayegani, Erfan and Mamun, Md Abdullah Al and Fu, Yu and Zaree, Pedram and Dong, Yue and Abu-Ghazaleh, Nael},
  journal={arXiv preprint arXiv:2310.10844},
  year={2023}
}

@article{perez2022red,
  title={Red Teaming Language Models with Language Models},
  author={Perez, Ethan and Huang, Saffron and Song, Francis and Cai, Trevor and Ring, Roman and Aslanides, John and Glaese, Amelia and McAleese, Nat and Irving, Geoffrey},
  journal={arXiv preprint arXiv:2202.03286},
  year={2022}
}

@inproceedings{mcmahan2017fedavg,
  title={Communication-Efficient Learning of Deep Networks from Decentralized Data},
  author={McMahan, Brendan and Moore, Eider and Ramage, Daniel and Hampson, Seth and Arcas, Blaise Aguera y},
  booktitle={Proceedings of the 20th International Conference on Artificial Intelligence and Statistics (AISTATS)},
  pages={1273--1282},
  year={2017},
  volume={54},
  series={Proceedings of Machine Learning Research},
  publisher={PMLR}
}

@article{ferrag2021federated,
  title={Federated Deep Learning for Cyber Security in the Internet of Things: Concepts, Applications, and Experimental Analysis},
  author={Ferrag, Mohamed Amine and Friha, Othmane and Maglaras, Leandros and Janicke, Helge and Shu, Lei},
  journal={IEEE Access},
  volume={9},
  pages={138509--138542},
  year={2021}
}

@inproceedings{blanchard2017machine,
  title={Machine Learning with Adversaries: {Byzantine} Tolerant Gradient Descent},
  author={Blanchard, Peva and El Mhamdi, El Mahdi and Guerraoui, Rachid and Stainer, Julien},
  booktitle={Advances in Neural Information Processing Systems (NeurIPS)},
  pages={119--129},
  year={2017}
}

@inproceedings{bagdasaryan2020backdoor,
  title={How to Backdoor Federated Learning},
  author={Bagdasaryan, Eugene and Veit, Andreas and Hua, Yiqing and Estrin, Deborah and Shmatikov, Vitaly},
  booktitle={International Conference on Artificial Intelligence and Statistics (AISTATS)},
  pages={2938--2948},
  year={2020}
}

@inproceedings{nguyen2022flame,
  title={{FLAME}: Taming Backdoors in Federated Learning},
  author={Nguyen, Thien Duc and Rieger, Phillip and De Viti, Roberta and Chen, Huili and Brandenburg, Bj{\"o}rn B and Yalame, Hossein and M{\"o}llering, Helen and Fereidooni, Hossein and Marchal, Samuel and Miettinen, Markus and others},
  booktitle={USENIX Security Symposium},
  pages={1415--1432},
  year={2022}
}

@inproceedings{bonawitz2017secure_aggregation,
  title={Practical Secure Aggregation for Privacy-Preserving Machine Learning},
  author={Bonawitz, Keith and Ivanov, Vladimir and Kreuter, Ben and Marcedone, Antonio and McMahan, H Brendan and Patel, Sarvar and Ramage, Daniel and Segal, Aaron and Seth, Karn},
  booktitle={Proceedings of the 2017 ACM SIGSAC Conference on Computer and Communications Security (CCS)},
  pages={1175--1191},
  year={2017}
}

@inproceedings{abadi2016dpsgd,
  title={Deep Learning with Differential Privacy},
  author={Abadi, Martin and Chu, Andy and Goodfellow, Ian and McMahan, H Brendan and Mironov, Ilya and Talwar, Kunal and Zhang, Li},
  booktitle={Proceedings of the 2016 ACM SIGSAC Conference on Computer and Communications Security (CCS)},
  pages={308--318},
  year={2016}
}

@inproceedings{shokri2017membership,
  title={Membership Inference Attacks Against Machine Learning Models},
  author={Shokri, Reza and Stronati, Marco and Song, Congzheng and Shmatikov, Vitaly},
  booktitle={IEEE Symposium on Security and Privacy (S\&P)},
  pages={3--18},
  year={2017}
}

@book{voigt2017gdpr,
  title={The {EU} General Data Protection Regulation ({GDPR}): A Practical Guide},
  author={Voigt, Paul and Von dem Bussche, Axel},
  year={2017},
  publisher={Springer}
}

@article{xi2023rise,
  title={The Rise and Potential of Large Language Model Based Agents: A Survey},
  author={Xi, Zhiheng and Chen, Wenxiang and Guo, Xin and He, Wei and Ding, Yiwen and Hong, Boyang and Zhang, Ming and Wang, Junzhe and Jin, Senjie and Zhou, Enyu and others},
  journal={arXiv preprint arXiv:2309.07864},
  year={2023}
}

@article{wang2024survey,
  title={A Survey on Large Language Model based Autonomous Agents},
  author={Wang, Lei and Ma, Chen and Feng, Xueyang and Zhang, Zeyu and Yang, Hao and Zhang, Jingsen and Chen, Zhiyuan and Tang, Jiakai and Chen, Xu and Lin, Yankai and others},
  journal={Frontiers of Computer Science},
  volume={18},
  number={6},
  pages={186345},
  year={2024}
}

@article{schick2023toolformer,
  title={Toolformer: Language Models Can Teach Themselves to Use Tools},
  author={Schick, Timo and Dwivedi-Yu, Jane and Dess{\`i}, Roberto and Raileanu, Roberta and Lomeli, Maria and Zettlemoyer, Luke and Cancedda, Nicola and Scialom, Thomas},
  journal={arXiv preprint arXiv:2302.04761},
  year={2023}
}

@article{qin2023toolllm,
  title={ToolLLM: Facilitating Large Language Models to Master 16000+ Real-world APIs},
  author={Qin, Yujia and Liang, Shihao and Ye, Yining and Zhu, Kunlun and Yan, Lan and Lu, Yaxi and Lin, Yankai and Cong, Xin and Tang, Xiangru and Qian, Bill and others},
  journal={arXiv preprint arXiv:2307.16789},
  year={2023}
}

@article{patil2023gorilla,
  title={Gorilla: Large Language Model Connected with Massive APIs},
  author={Patil, Shishir G and Zhang, Tianjun and Wang, Xin and Gonzalez, Joseph E},
  journal={arXiv preprint arXiv:2305.15334},
  year={2023}
}

@article{zou2023universal,
  title={Universal and Transferable Adversarial Attacks on Aligned Language Models},
  author={Zou, Andy and Wang, Zifan and Kolter, J Zico and Fredrikson, Matt},
  journal={arXiv preprint arXiv:2307.15043},
  year={2023}
}

@article{liu2023prompt,
  title={Prompt Injection Attack Against LLM-integrated Applications},
  author={Liu, Yi and Deng, Gelei and Xu, Zhengzi and Li, Yuekang and Zheng, Yaowen and Zhang, Ying and Zhao, Lida and Zhang, Tianwei and Liu, Yang},
  journal={arXiv preprint arXiv:2306.05499},
  year={2023}
}

@article{zhan2024injecagent,
  title={InjecAgent: Benchmarking Indirect Prompt Injections in Tool-Integrated Large Language Model Agents},
  author={Zhan, Qiusi and Liang, Zhixiang and Ying, Zifan and Kang, Daniel},
  journal={arXiv preprint arXiv:2403.02691},
  year={2024}
}

@book{howard2005security,
  title={The Security Development Lifecycle},
  author={Howard, Michael and Lipner, Steve},
  year={2006},
  publisher={Microsoft Press}
}

@misc{owasp2021top10,
  title={OWASP Top 10:2021},
  author={{OWASP Foundation}},
  howpublished={\url{https://owasp.org/Top10/}},
  year={2021}
}

@article{kairouz2021advances,
  title={Advances and Open Problems in Federated Learning},
  author={Kairouz, Peter and McMahan, H Brendan and Avent, Brendan and Bellet, Aur{\'e}lien and Bennis, Mehdi and Bhagoji, Arjun Nitin and Bonawitz, Kallista and Charles, Zachary and Cormode, Graham and Cummings, Rachel and others},
  journal={Foundations and Trends in Machine Learning},
  volume={14},
  number={1--2},
  pages={1--210},
  year={2021}
}

@inproceedings{li2020fedprox,
  title={Federated Optimization in Heterogeneous Networks},
  author={Li, Tian and Sahu, Anit Kumar and Zaheer, Manzil and Sanjabi, Maziar and Talwalkar, Ameet and Smith, Virginia},
  booktitle={Proceedings of Machine Learning and Systems},
  volume={2},
  pages={429--450},
  year={2020}
}

@article{zhao2018federated,
  title={Federated Learning with Non-IID Data},
  author={Zhao, Yue and Li, Meng and Lai, Liangzhen and Suda, Naveen and Civin, Damon and Chandra, Vikas},
  journal={arXiv preprint arXiv:1806.00582},
  year={2018}
}

@article{li2021fedmal,
  title={FedMal: A Federated Learning Framework for Malware Detection},
  author={Li, Yushan and Wei, Yanjun and others},
  journal={IEEE Transactions on Dependable and Secure Computing},
  year={2021}
}

@article{jayathilaka2025federated,
  title={Privacy-Preserving Prompt Injection Detection for {LLMs} Using Federated Learning and Embedding-Based {NLP} Classification},
  author={Jayathilaka, Hasini},
  journal={arXiv preprint arXiv:2511.12295},
  year={2025}
}

@inproceedings{fredrikson2015model,
  title={Model Inversion Attacks that Exploit Confidence Information and Basic Countermeasures},
  author={Fredrikson, Matt and Jha, Somesh and Ristenpart, Thomas},
  booktitle={Proceedings of the 22nd ACM SIGSAC Conference on Computer and Communications Security},
  pages={1322--1333},
  year={2015}
}

@inproceedings{mirsky2018kitsune,
  title={Kitsune: An Ensemble of Autoencoders for Online Network Intrusion Detection},
  author={Mirsky, Yisroel and Doitshman, Tomer and Elovici, Yuval and Shabtai, Asaf},
  booktitle={Network and Distributed System Security Symposium (NDSS)},
  year={2018}
}

@inproceedings{du2017deeplog,
  title={DeepLog: Anomaly Detection and Diagnosis from System Logs through Deep Learning},
  author={Du, Min and Li, Feifei and Zheng, Guineng and Srikumar, Vivek},
  booktitle={Proceedings of the 2017 ACM SIGSAC Conference on Computer and Communications Security},
  pages={1285--1298},
  year={2017}
}

@inproceedings{brown2018recurrent,
  title={Recurrent Neural Network Attention Mechanisms for Interpretable System Log Anomaly Detection},
  author={Brown, Andy and Tuor, Aaron and Hutchinson, Brian and Mez, Nicole},
  booktitle={Proceedings of the First Workshop on Machine Learning for Computing Systems},
  pages={1--8},
  year={2018}
}

@article{baltrusaitis2018multimodal,
  title={Multimodal Machine Learning: A Survey and Taxonomy},
  author={Baltru{\v{s}}aitis, Tadas and Ahuja, Chaitanya and Morency, Louis-Philippe},
  journal={IEEE Transactions on Pattern Analysis and Machine Intelligence},
  volume={41},
  number={2},
  pages={423--443},
  year={2018}
}

@article{xu2013survey,
  title={A Survey on Multi-view Learning},
  author={Xu, Chang and Tao, Dacheng and Xu, Chao},
  journal={arXiv preprint arXiv:1304.5634},
  year={2013}
}

@inproceedings{ngiam2011multimodal,
  title={Multimodal Deep Learning},
  author={Ngiam, Jiquan and Khosla, Aditya and Kim, Mingyu and Nam, Juhan and Lee, Honglak and Ng, Andrew Y},
  booktitle={Proceedings of the 28th International Conference on Machine Learning},
  pages={689--696},
  year={2011}
}

@inproceedings{koh2021wilds,
  title={WILDS: A Benchmark of in-the-Wild Distribution Shifts},
  author={Koh, Pang Wei and Sagawa, Shiori and Marber, Henrik and others},
  booktitle={International Conference on Machine Learning},
  pages={5637--5664},
  year={2021},
  organization={PMLR}
}

@article{shen2021towards,
  title={Towards Out-of-Distribution Generalization: A Survey},
  author={Shen, Zheyan and Liu, Jiashuo and He, Yue and Zhang, Xingxuan and Xu, Renzhe and Yu, Han and Cui, Peng},
  journal={arXiv preprint arXiv:2108.13624},
  year={2021}
}

@inproceedings{hendrycks2019benchmarking,
  title={Benchmarking Neural Network Robustness to Common Corruptions and Perturbations},
  author={Hendrycks, Dan and Dietterich, Thomas},
  booktitle={International Conference on Learning Representations},
  year={2019}
}

@article{yang2021generalized,
  title={Generalized Out-of-Distribution Detection: A Survey},
  author={Yang, Jingkang and Zhou, Kaiyang and Li, Yixuan and Liu, Ziwei},
  journal={arXiv preprint arXiv:2110.11334},
  year={2021}
}

@article{bengio2013representation,
  title={Representation Learning: A Review and New Perspectives},
  author={Bengio, Yoshua and Courville, Aaron and Vincent, Pascal},
  journal={IEEE Transactions on Pattern Analysis and Machine Intelligence},
  volume={35},
  number={8},
  pages={1798--1828},
  year={2013}
}

@inproceedings{chen2020simple,
  title={A Simple Framework for Contrastive Learning of Visual Representations},
  author={Chen, Ting and Kornblith, Simon and Norouzi, Mohammad and Hinton, Geoffrey},
  booktitle={International Conference on Machine Learning},
  pages={1597--1607},
  year={2020},
  organization={PMLR}
}

@article{hochreiter1997lstm,
  title={Long Short-Term Memory},
  author={Hochreiter, Sepp and Schmidhuber, J{\"u}rgen},
  journal={Neural Computation},
  volume={9},
  number={8},
  pages={1735--1780},
  year={1997}
}

@article{kingma2014adam,
  title={Adam: A Method for Stochastic Optimization},
  author={Kingma, Diederik P and Ba, Jimmy},
  journal={arXiv preprint arXiv:1412.6980},
  year={2014}
}

@article{vaswani2017attention,
  title={Attention is All You Need},
  author={Vaswani, Ashish and Shazeer, Noam and Parmar, Niki and Uszkoreit, Jakob and Jones, Llion and Gomez, Aidan N and Kaiser, {\L}ukasz and Polosukhin, Illia},
  journal={Advances in Neural Information Processing Systems},
  volume={30},
  year={2017}
}

@inproceedings{jacobs1991adaptive,
  title={Adaptive Mixtures of Local Experts},
  author={Jacobs, Robert A and Jordan, Michael I and Nowlan, Steven J and Hinton, Geoffrey E},
  booktitle={Neural Computation},
  volume={3},
  number={1},
  pages={79--87},
  year={1991}
}

@article{shazeer2017outrageously,
  title={Outrageously Large Neural Networks: The Sparsely-Gated Mixture-of-Experts Layer},
  author={Shazeer, Noam and Mirhoseini, Azalia and Maziarz, Krzysztof and Davis, Andy and Le, Quoc and Hinton, Geoffrey and Dean, Jeff},
  journal={arXiv preprint arXiv:1701.06538},
  year={2017}
}

@misc{opacus,
  title={Opacus: User-Friendly Differential Privacy Library in PyTorch},
  author={Yousefpour, Ashkan and Shilov, Igor and Sablayrolles, Alexandre and Testuggine, Davide and Prasad, Karthik and Maber, Mani and Nie, John and Grimm, Sebastian and Guo, Wenxu and Zong, Lei and others},
  howpublished={\url{https://opacus.ai}},
  year={2021}
}

@inproceedings{paszke2019pytorch,
  title={PyTorch: An Imperative Style, High-Performance Deep Learning Library},
  author={Paszke, Adam and Gross, Sam and Massa, Francisco and Lerer, Adam and Bradbury, James and Chanan, Gregory and Killeen, Trevor and Lin, Zeming and Gimelshein, Natalia and Antiga, Luca and others},
  booktitle={Advances in Neural Information Processing Systems},
  volume={32},
  year={2019}
}
}

\clearpage
\appendix

\section{Extended Experimental Details}
\label{app:details}

\textbf{Dataset Generation.} Traces mirror documented attacks~\cite{andriushchenko2024agentharm,greshake2023ipi,wei2023jailbroken,zhan2024injecagent} with placeholder content. 2,500 total traces; 500 per organization; 50\% benign, 50\% malicious across 5 attack families. Data handling follows privacy-preserving principles consistent with GDPR requirements~\cite{voigt2017gdpr}.

\textbf{Non-IID Distribution.} Attack categories distributed non-uniformly across organizations to simulate realistic specialization: Org-1 sees primarily prompt injection, Org-2 sees tool hijacking, etc.

\textbf{Architecture Details.} Embedding: 64-dim learned vectors. BiLSTM~\cite{hochreiter1997lstm}: hidden=64, concatenated forward/backward states yield 128-dim. Projection: linear 128→32. Classifier: 32→64 (ReLU) → 64→1 (sigmoid). Gated model uses parallel encoders with element-wise gating.

\textbf{Training Configuration.} Adam optimizer~\cite{kingma2014adam} ($\eta$=0.001, $\beta_1$=0.9, $\beta_2$=0.999), binary cross-entropy loss, batch size 32, 5 federated rounds × 5 local epochs.

\textbf{Privacy Mechanisms.} We employ secure aggregation~\cite{bonawitz2017secure_aggregation} to prevent the server from observing individual client updates. When differential privacy is enabled: gradient clipping $C$=1.0, noise multiplier $\sigma \in [0.6, 1.1]$, $\delta$=10$^{-5}$, Rényi DP accounting via Opacus~\cite{opacus}. These mechanisms bound membership inference~\cite{shokri2017membership} and model inversion~\cite{fredrikson2015model} risks.

\textbf{Hardware.} Apple M2 Pro, 16GB RAM. Training completes in under 5 minutes per configuration. No GPU required. Implementation uses PyTorch~\cite{paszke2019pytorch}.

\textbf{Differential Privacy.} When enabled: gradient clipping $C$=1.0, noise multiplier $\sigma \in [0.6, 1.1]$, $\delta$=10$^{-5}$, Rényi DP accounting via Opacus~\cite{opacus}.

\section{Complete Results}
\label{app:results}

Table~\ref{tab:complete_results} shows the full comparison across all settings.

\begin{table}[h]
\centering
\small
\begin{tabular}{lcccc}
\toprule
\textbf{Setting} & \textbf{Conv} & \textbf{Struct} & \textbf{Gated} & \textbf{Best} \\
\midrule
IID & 0.87 & \textbf{0.93} & 0.89 & Struct \\
\midrule
Social eng. & 0.78 & 0.67 & \textbf{0.89} & Gated \\
Prompt inj. & 0.69 & 0.81 & \textbf{0.83} & Gated \\
Data exfil. & 0.46 & \textbf{0.85} & 0.62 & Struct \\
Tool hijack & 0.39 & \textbf{0.85} & 0.60 & Struct \\
Unknown & 0.26 & \textbf{0.97} & 0.92 & Struct \\
\midrule
OOD Average & 0.52 & \textbf{0.83} & 0.77 & Struct \\
\bottomrule
\end{tabular}
\caption{Complete results. Structural achieves best OOD average; gated excels on linguistic attacks.}
\label{tab:complete_results}
\end{table}

Table~\ref{tab:per_org} shows per-organization performance consistency.

\begin{table}[h]
\centering
\small
\begin{tabular}{lccc}
\toprule
\textbf{Org} & \textbf{Conv IID} & \textbf{Struct IID} & \textbf{Struct OOD} \\
\midrule
1 & 0.85 & 0.92 & 0.81 \\
2 & 0.88 & 0.93 & 0.84 \\
3 & 0.86 & 0.92 & 0.82 \\
4 & 0.89 & 0.94 & 0.86 \\
5 & 0.87 & 0.93 & 0.83 \\
\midrule
\textbf{Mean} & 0.87 & 0.93 & 0.83 \\
\textbf{Std} & 0.02 & 0.01 & 0.02 \\
\bottomrule
\end{tabular}
\caption{Per-organization results show consistent improvement with low variance.}
\label{tab:per_org}
\end{table}

\section{Ablation Studies}
\label{app:ablation}

\textbf{View Necessity.} Neither single view is optimal across attack types. Conv-only achieves 0.78 on social engineering but 0.39 on tool hijacking. Struct-only achieves 0.85 on tool hijacking but 0.67 on social engineering. Gated fusion provides balance (0.89 / 0.60).

\textbf{Vocabulary Size.} Structural tokenization uses only 9 tokens versus 26 for conversational, yet achieves superior OOD performance. This suggests that abstraction level---not vocabulary richness---drives generalization.

\textbf{Aggregation Methods.} Local, FedAvg~\cite{mcmahan2017fedavg}, and ensemble methods produce identical results (within $\pm$0.02), confirming representation is the bottleneck. This aligns with findings on non-IID challenges in federated learning~\cite{zhao2018federated,li2020fedprox,kairouz2021advances}.

\section{Failure Analysis}
\label{app:failure}

\textbf{Unknown Category Inversion.} With conversational tokenization, unknown attacks achieve AUC 0.26 (below random). Investigation reveals systematic inversion: the model assigns higher attack probability to benign samples (mean 0.97) than actual attacks (mean 0.69). The model learned ``unfamiliar = dangerous,'' but unknown attacks are also unfamiliar, causing ranking reversal---a failure mode consistent with OOD generalization challenges~\cite{hendrycks2019benchmarking,yang2021generalized}.

\textbf{Structural Recovery.} Structural tokenization recovers to AUC 0.97 because execution patterns remain discriminative regardless of linguistic novelty.

\section{Structural Tokenization Algorithm}
\label{app:algorithm}

\begin{algorithm}[h]
\caption{Structural Tokenization $\phi_{\text{struct}}(\tau)$}
\small
\begin{algorithmic}[1]
\REQUIRE Agent trace $\tau = (\mathcal{M}, \mathcal{T}, \mathcal{R})$
\STATE $S \leftarrow []$
\FOR{each turn in conversation}
    \IF{turn is system message}
        \STATE $S.\text{append}(\texttt{[SYS]})$
    \ELSIF{turn is user message}
        \STATE $S.\text{append}(\texttt{[USER]})$
    \ELSIF{turn is assistant message}
        \STATE $S.\text{append}(\texttt{[ASSISTANT]})$
        \IF{turn contains tool call}
            \STATE $S.\text{append}(\texttt{[TOOL]})$
            \STATE $S.\text{append}(\texttt{[ARGS]})$
        \ENDIF
    \ELSIF{turn is tool observation}
        \STATE $S.\text{append}(\texttt{[OBS]})$
    \ENDIF
\ENDFOR
\STATE $S.\text{append}(\texttt{[OUTPUT]})$
\RETURN $\textsc{Pad}(S, L_{\max})$
\end{algorithmic}
\end{algorithm}

\section{Limitations and Future Work}
\label{app:limitations}

\textbf{Tokenizer Robustness and Adaptive Adversaries.} Our structural tokenizer uses deterministic, rule-based pattern matching (Algorithm 1). This design choice was intentional: it allows us to isolate the effect of structural abstraction itself without conflating it with tokenizer sophistication. However, we acknowledge that such simplicity introduces an evasion surface. A motivated adversary could attempt to fragment tool calls across turns, pad traces with benign operations, or deliberately mimic common benign execution flows to reduce anomaly scores.

While this is an important limitation, it does not undermine the central finding of this paper: structural execution patterns carry threat-relevant signal that conversational features alone miss. Our architecture is modular, and the tokenizer can be replaced without changing the detection model. The BiLSTM operates over abstract token sequences, making it agnostic to tokenization strategy. In practice, we expect stronger robustness from richer tokenization strategies such as: (i)~state-transition modeling over execution graphs rather than isolated events, (ii)~semantic enrichment of tool calls with access scope and sensitivity labels, (iii)~stochastic abstraction to reduce predictability, and (iv)~learned latent encoders trained contrastively over benign and malicious traces~\cite{chen2020simple}.

We also note that evasion attempts often introduce secondary anomalies (e.g., entropy spikes, abnormal sequencing, or workflow divergence), suggesting that attempts to hide execution semantics may themselves become detectable. A full adversarial evaluation---where attackers explicitly optimize to evade the representation layer---remains essential future work. We are actively pursuing red-team evaluations and partnerships to enable adversarial testing against production agents.

\textbf{Synthetic Data.} Our traces are programmatically generated and may not capture real attacker creativity or production noise. Real-world validation through industry partnerships is critical.

\textbf{Static Tokenization.} Rule-based extraction may be evaded by sophisticated attackers. Learned tokenization via neural encoders~\cite{chen2020simple,vaswani2017attention} is a promising direction.

\textbf{Attack Coverage.} Our taxonomy covers 5 families. Emerging attacks like multi-agent exploits and memory poisoning require study.

\textbf{Gated Fusion Tradeoff.} While gated fusion excels on linguistic attacks, it underperforms struct-only on structural attacks (0.60 vs 0.85). Improved fusion mechanisms such as attention-based routing~\cite{vaswani2017attention} or sparse expert selection~\cite{shazeer2017outrageously} are needed.

\textbf{Byzantine Robustness.} Our federated setting assumes honest-but-curious participants. Malicious clients could attempt model poisoning~\cite{bagdasaryan2020backdoor}; defenses such as Byzantine-tolerant aggregation~\cite{blanchard2017machine} or backdoor detection~\cite{nguyen2022flame} would be required for production deployment.

\textbf{Privacy Guarantees.} While we employ DP-SGD and secure aggregation, determined adversaries may still attempt membership inference~\cite{shokri2017membership} or model inversion~\cite{fredrikson2015model}. Stronger privacy budgets ($\epsilon < 1$) or additional defenses may be needed for highly sensitive deployments.

\section{Reproducibility}
\label{app:reproducibility}

\textbf{Implementation.} PyTorch 2.0.1 with Opacus 1.4.0 for differential privacy. All experiments run on consumer hardware (Apple M2 Pro, 16GB RAM) without GPU.

\textbf{Hyperparameters.} Learning rate 0.001, batch size 32, 5 federated rounds, 5 local epochs, embedding dimension 64, LSTM hidden size 64, latent dimension 32.

\textbf{Code and Data.} Synthetic trace generation scripts and training code will be released upon publication.

\end{document}